\documentclass[aps,twocolumn,superscriptaddress,floatfix,preprintnumbers,nofootinbib, nobibnotes]{revtex4-1}
\usepackage[utf8]{inputenc}
\usepackage{amsmath}
\usepackage{amssymb}

\usepackage{enumerate}
\usepackage{amsmath}
\usepackage{tikz}
\usepackage{feynmf}
\usepackage{slashed}
\usepackage{braket}
\usepackage[toc,page]{appendix}
\usepackage{url}
\usepackage{graphicx}
\usepackage[pdftex,bookmarks,linktocpage,pdfpagelabels,plainpages=false,hyperfigures,linkcolor=blue,citecolor=blue]{hyperref}

\hypersetup{colorlinks=true}

\usepackage{mathrsfs}  
\usepackage{cancel}
\usepackage[normalem]{ulem}
\usepackage[perpage]{footmisc}
\usepackage[capitalize]{cleveref}

\newcommand{\beq}{\begin{equation}}
\newcommand{\eeq}{\end{equation}}

\usepackage{hyperref}
\definecolor{nicered}{rgb}{0.7,0.1,0.1}
\definecolor{nicegreen}{RGB}{53,170,102}
\definecolor{niceblue}{rgb}{0.117,0.5625,1.0}
\definecolor{nicepurple}{RGB}{127, 38, 222}
\hypersetup{colorlinks,
			citecolor      = nicepurple, 
			linkcolor       = niceblue, 
			urlcolor        = nicegreen, 
			anchorcolor = niceblue}

\begin{document}


{\hfill IFT-UAM/CSIC-23-145, MI-HET-820}

\title{Heavy Neutral Leptons via Axion-Like Particles at Neutrino Facilities}

\author{Asli~M~Abdullahi}
\affiliation{Instituto de F\'{i}sica Te\'{o}rica UAM-CSIC, Universidad Aut\'{o}noma de Madrid, Cantoblanco, 28049 Madrid, Spain}
\affiliation{
Particle Theory Department, Fermi National Accelerator Laboratory, Batavia, IL, 60510, USA}
\author{Andr\'{e} de Gouv\^{e}a} 
\affiliation{Northwestern University, Department of Physics \& Astronomy, 2145 Sheridan Road, Evanston, IL 60208, USA}
\author{Bhaskar~Dutta}
\affiliation{Mitchell Institute for Fundamental Physics and Astronomy,
Department of Physics and Astronomy, Texas A\&M University, College Station, TX 77843, USA}
\author{Ian~M.~Shoemaker}
\affiliation{Center for Neutrino Physics, Department of Physics, Virginia Tech, Blacksburg, VA 24061, USA}
\author{Zahra~Tabrizi}
\affiliation{Northwestern University, Department of Physics \& Astronomy, 2145 Sheridan Road, Evanston, IL 60208, USA}

\begin{abstract}
Heavy neutral leptons (HNLs) are often among the hypothetical ingredients behind nonzero neutrino masses. If sufficiently light, they can be produced and detected in fixed-target-like experiments. We show that if the HNLs belong to a richer -- but rather generic -- dark sector, their production mechanism can deviate dramatically from expectations associated to the standard-model weak interactions. In more detail, we postulate that the dark sector contains an axion-like particle (ALP) that naturally decays into HNLs. Since ALPs mix with the pseudoscalar hadrons, the HNL flux might be predominantly associated to the production of neutral mesons (e.g., $\pi^0$, $\eta$) as opposed to charge hadrons (e.g., $\pi^\pm$, $K^\pm$). In this case, the physics responsible for HNL production and decay are not directly related and experiments like DUNE might be sensitive to HNLs that are too weakly coupled to the standard model to be produced via weak interactions, as is generically the case of HNLs that play a direct role in the type-I seesaw mechanism.  

\end{abstract}


\maketitle

{\bf Introduction.} 
\label{sec:Intro}
\setcounter{equation}{0}
The discovery of neutrino oscillation~\cite{Fukuda:1998mi,Fukuda:1998ah,Ahmad:2002jz}, and the confirmation of non-zero neutrino masses, implies the existence of fields beyond those of the standard model (SM) of particle physics. While little is known about these new degrees of freedom, a plethora of extensions have been proposed to explain the origin of the neutrino mass. Many of these extensions invoke the existence of SM-singlet fermions, commonly referred to as right-handed, or sterile, neutrinos. Popular examples of such models include the Type-1 seesaw mechanism~\cite{Minkowski:1977sc,Mohapatra:1979ia,GellMann:1980vs,Yanagida:1979as,Lazarides:1980nt,Mohapatra:1980yp,Schechter:1980gr,Cheng:1980qt,Foot:1988aq} and other seesaw variants~\cite{Schechter:1980gr, Konetschny:1977bn, Cheng:1980qt, Mohapatra:1980yp, Foot:1988aq, Ma:1998dn, Bajc:2006ia, Dorsner:2006fx}.
Depending on their properties, these singlet fermions may help shed light on other outstanding problems in particle physics, including the abundance of dark matter~\cite{Boehm:2003hm,Boehm:2003ha,Pospelov:2007mp,Pospelov:2008zw} and the baryon asymmetry of the universe~\cite{Fukugita:1986hr,Davidson:2008bu}.

SM-singlet fermions can interact with SM particles through Yukawa interactions involving the Higgs-doublet $H$ and the lepton-doublets $L$, in what is known as the neutrino portal. Such a coupling induces a mixing between the active neutrinos, $\nu_{e,\mu,\tau}$, and the SM-singlet fermions, $N_{s_j}$, $j=1,\ldots,n_s$, generating a Dirac mass for the neutrinos after electroweak symmetry breaking (EWSB). The singlet fermions may or may not have a Majorana mass, although such a mass term is always permitted by symmetry. We are left with mass eigenstates $\nu_i$, with masses $m_i$, that are linear superpositions of the active and singlet states, $\nu_{i} = U^{\dagger}_{\alpha i}\nu_\alpha$, where $\alpha$ now runs over $\alpha=e,\mu,\tau,s_1, s_2,\ldots, s_{n_s}$. The transformation is parametrized by the unitary $(3+n_s)\times(3+n_s)$ matrix, with components $U_{\alpha i}$. Henceforth, we will restrict our discussions to $n_s=1$ and refer to the new fermion as $N$. We are interested in mass scales of order 1~MeV to 1~GeV for the new particle, where $N$ is often referred to as a heavy neutral lepton (HNL). In this region of parameter space, the $U_{\alpha 4}$ elements of the mixing matrix often govern the production of $N$ in fixed-target experiments, along with their scattering cross sections inside detectors, and their decay properties.


The existence of SM-singet fermions invites one to consider a richer ``dark sector'' with its own particle content and interactions~\cite{Pospelov:2011ha,Harnik:2012ni,Batell:2016zod,Farzan:2016wym,DeRomeri:2017oxa,Magill:2018jla,Bertuzzo:2018ftf,Bertuzzo:2018itn,Ballett:2019cqp,Ballett:2019pyw,Coloma:2019qqj,Fischer:2019fbw,Cline:2020mdt,Berbig:2020wve}. Generically, it is useful to allow for the possibility that $N$, while a SM-singlet, is charged under a ``dark'' gauge group and interacts with other particles that are also SM-singlets. 

We will concentrate on dark sectors that also contain a new pseudoscalar, $a$, with non-zero mass, sometimes referred to as an axion-like particle (ALP). We will further assume that $a$ couples to $N$ and that its mass is such that it decays predominantly to pairs of HNL. This is not an especially extraordinary hypothesis as it mimics the SM. In the SM, the strong interactions confine and the lightest propagating hadronic degrees of freedom are pseudoscalars (pions, kaons, etc). The lightest among these are restricted to decaying weakly into pairs of leptons (e.g., $\pi^+\to\mu^+\nu$) or electromagnetically into photons (e.g., $\pi^0\to\gamma\gamma$). If the dark sector is anything like the SM, it is easy to imagine scenarios where $a\to NN$ is similarly inevitable. A complete chiral dark sector model with these characteristics was proposed and explored in  Ref.~\cite{Berryman:2017twh}. There, the ALPs were referred to as `dark pions', and the HNLs `dark neutrinos'. We will refer to Ref.~\cite{Berryman:2017twh} when discussing concrete realizations of the scenario of interest, but emphasize that our results remain general and not tied to any specific model. 


In the absence of the ALP, in fixed-target-like experiments and for HNL masses below a GeV, the HNLs are typically produced in the decay of charged mesons (e.g., $\pi^+\rightarrow\mu^+N$). In this case, the production rate is proportional to $|U_{\alpha 4}|^2$. Further downstream, the $N$ can be detected via its decays to SM particles, with the $N$ partial widths also proportional to $|U_{\alpha 4}|^2$. If a kinematically accessible ALP is now added to the picture, we allow for a new HNL production mechanism in fixed-target-like experiments. The ALPs mix with {\sl neutral} pseudoscalar mesons ($\pi^0,\eta$, etc.) and are produced in tandem with light hadrons at the target. This production rate is, of course, independent of $U_{\alpha 4}$. The ALPs then decay to HNL pairs with large branching ratios and HNL decays downstream can lead to an observable signal in the detector. 

\begin{figure}[ht]
    \centering
\includegraphics[width=0.47\textwidth]{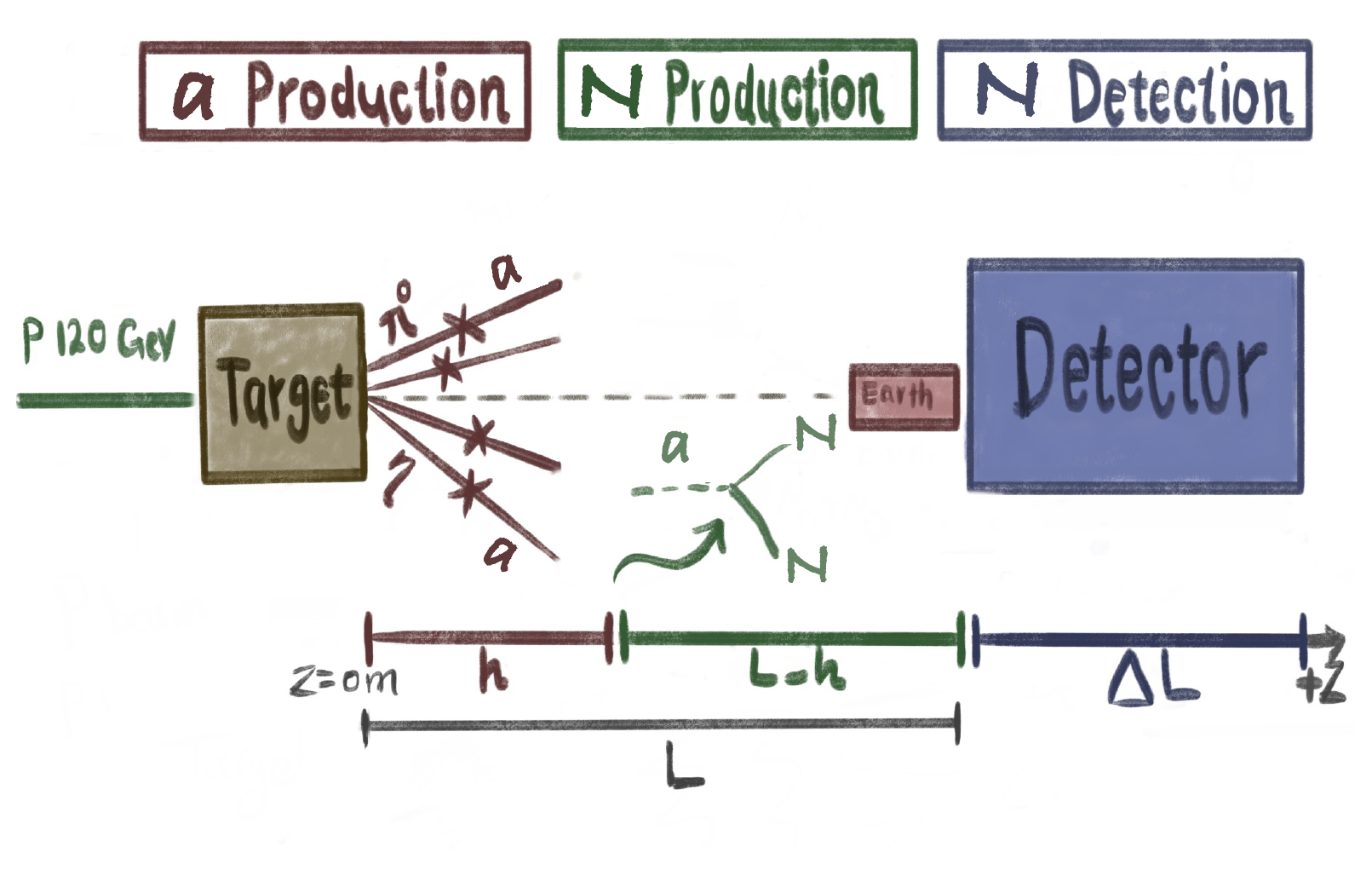}
\caption{
  Schematic description of the production of axion-like particles (ALP, $a$) followed by the production of heavy neutral leptons (HNL, $N$), in a fixed-target experimental setup. In the DUNE near-detector complex, a beam of 120~GeV protons impinges on a target, producing charged and neutral mesons. ALP production is proportional to that of neutral mesons: the crosses represent $a$--neutral-meson mixing. The ALP travels towards the detector and decays into a pair of HNLs a distance $h$ from the target. The HNLs that reach the detector (width $\Delta L$, a distance $L$ from the target) may decay inside the detector volume into SM particles, yielding an observable signal. Not to scale.
  }
\label{fig:Schematic}
\end{figure}

While generic ALP-HNL couplings have been studied in the literature, the focus has typically been on the constraint posed by experiment on the coupling itself~\cite{deGiorgi:2022oks, Alves:2019xpc}, or the contribution to existing processes coming from such a coupling~\cite{Bonilla:2023dtf}. In this work, we focus on the impact of an ALP-HNL coupling on the discovery potential of HNLs in fixed-target-like experiments. We explore the consequences of this new HNL production mechanism at the DUNE near-detector complex, taken as a particularly relevant experimental setup. Fig.~\ref{fig:Schematic} provides a schematic of the production--decay--detection process. We find that, depending on the ALP properties, one can explore previously inaccessible regions of the HNL parameter space, including the region preferred by the type-I seesaw model for neutrino masses. 
{\bf Heavy ALP production.} 
\label{sec:DarkPion}
%
The ALP $a$ is a pseudo-scalar field and mixes with the SM pseudoscalar mesons, including the $\pi^0$ and the $\eta$. Following Refs.~\cite{Krauss:1986bq,Alves:2017avw,Altmannshofer:2019yji} and especially Ref.~\cite{Kelly:2020dda}, the relevant mixing contributions of the SM mesons to the ALP can be defined as 
%
\begin{align}
\pi^0&\rightarrow\pi^0+g_{\pi a}a\,,\\
\eta&\rightarrow\eta^0+g_{\eta a}a\,,\label{eq:piphys}
\end{align}
where the meson mixing parameters are \cite{Kelly:2020dda}
\begin{align}
g_{\pi a}&=\frac{1}{6}\frac{f_\pi}{f_{a}}\frac{m_{a}^2}{m_{a}^2-m_{\pi^0}^2}\,,\label{eq:gpi0}\\
g_{\eta a}&=\frac{1}{\sqrt{6}}\frac{f_\pi}{f_{a}}\bigg(\frac{m_{a}^2-\frac{4}{9}m_{\pi^0}^2}{m_{a}^2-m_{\eta}^2}\bigg)\,.\label{eq:geta}
\end{align}

In order to estimate the ALP production rate at a fixed-target facility, we assume that the probability of producing an $a$ particle with three-momentum $\vec{p}_a$ is the same as that to produce a pseudoscalar $\mathbf{m}$ with the same three-momentum, up to a scaling factor $g_{\mathbf{m} a}^2$. This is a good approximation as long as the mass of the ALP is similar to the hadronic pseudoscalar masses. With this in mind, the ALP differential flux is proportional to that of the pseudoscalar $\mathbf{m}$:
%
%
%
\begin{align}\label{eq:diffflus}
\frac{d^2\phi_{a}}{dE_{a}d\theta_a}=g_{\mathbf{m}a}^2\frac{E_{a}}{E_{\mathbf{m}}}\frac{d^2\phi_{\mathbf{m}}}{dE_{\mathbf{m}}d\theta_\mathbf{m}}\,,
\end{align}
where $\phi_{a,\mathbf{m}}$ are the ALP and meson fluxes, $E_{a,\mathbf{m}}$ their respective energies, and $\theta_{a}=\theta_\mathbf{m}$ is the angle defined by the outgoing ALP/meson three-momentum and the direction of the incoming proton beam. For meson fluxes at the DUNE near-detector facility, assuming the SM, we used the results from Refs.~\cite{Brdar:2020dpr,Brdar:2022vum}, obtained with GEANT4~\cite{Agostinelli:2002hh}

\begin{figure}[ht]
    \centering
\includegraphics[width=0.47\textwidth]{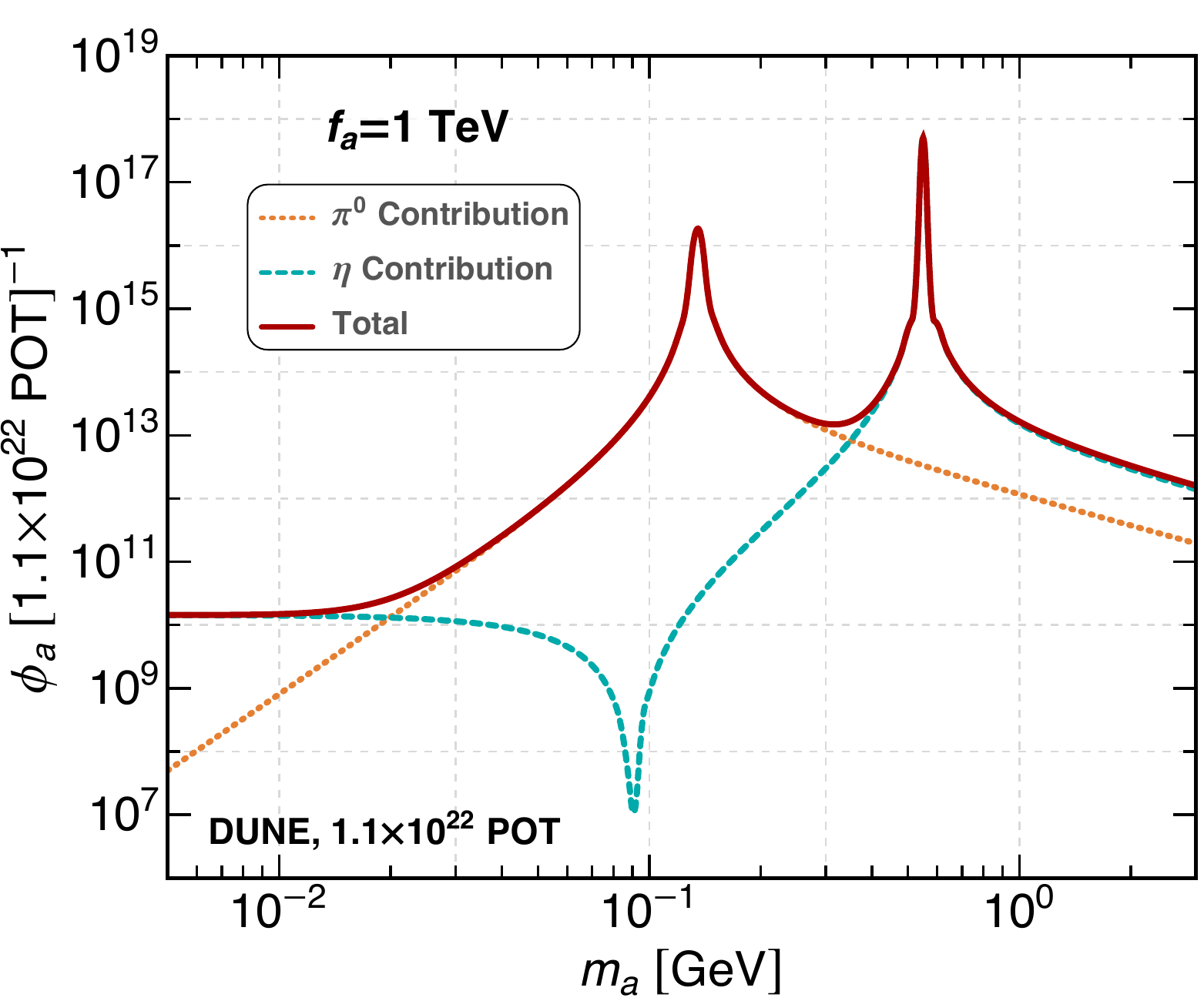}
\caption{
 Expected ALP flux at the DUNE near-detector facility as a function of the ALP mass $m_{a}$. The dotted, orange (dashed, blue) curve corresponds to the contribution from ALP--$\pi^0$ (ALP--$\eta$) mixing while the total flux is depicted with the solid, red curve. We assume a total exposure of $1.1\times10^{22}$ protons on target and $f_{a}=1$~TeV. The ALP flux is proportional to $f_{a}^{-2}$), see text for details.}
\label{fig:DarkPionFlux}
\end{figure}

Integrating Eq.~(\ref{eq:diffflus}) over $\theta_a$ and $E_a$, Fig.~\ref{fig:DarkPionFlux} depicts the ALP flux as a function of the ALP mass $m_a$, considering the contributions from ALP--$\pi^0$ (dotted, orange curve) and ALP--$\eta$ (dashed, blue curve) mixing. We assume a proton-on-target (POT) rate of $1.1 \times 10^{21}$ per year for a period of 10 years~\cite{DUNE:2020ypp}, and chose $f_{a}=1$~TeV. The ALP flux is proportional to $f_{a}^{-2}$ (see Eqs.~(\ref{eq:gpi0}), (\ref{eq:geta}) and (\ref{eq:diffflus})). We restrict the plot to $m_{a}$ values under a few GeV since the mixing formalism discussed above fails for larger values of $m_a$.

{\bf HNL production.} 
\label{sec:DarkN}
Once produced, we assume that the ALP decays exclusively into a pair of HNLs. For the decay rate, we use the model discussed in Ref.~\cite{Berryman:2017twh}, where ALP decays into HNLs parallel the decays of charged-pions into lepton pairs, mediated by $W$-boson exchange.
The partial width is given by,
\begin{align}   
\Gamma(a\to2N)=\frac{1}{4\pi v_D^4}f_{a}^2m_{a}m_{N}^2.
\end{align}
Here, $v_D$ is the vacuum expectation value of the dark sector Higgs-like scalar and $m_{N}$ is the HNL mass. Here we assume $v_D$ to be much larger than the weak scale \cite{Berryman:2017twh}.  Like SM pion decay, $a\to 2N$ is chirality suppressed. For very small $m_N$ values, the four-body decay, $a\to 4N$ is significant or even dominant (one SM parallel to this decay mode is $\pi^0\to Z^*Z^*\to 4\nu$). See Ref.~\cite{Berryman:2017twh} for details. Here, we restrict our discussion to regions of parameter space where the two-body final state dominates.

%
%
The probability that the ALP decays inside an infinitesimal interval $\Delta h$ a distance $h$ from the production target is 
\begin{align}
\left(\frac{{\rm d}P^{a}_{\rm{decay}}}{{\rm d}h}\right)\Delta h= \left(\frac{\Gamma}{(\beta\gamma)_a}e^{-\Gamma_a h/(\beta\gamma)_a}\right)\Delta h\,,
\end{align}
where $\Gamma_{a}$ is the ALP decay width and $(\beta\gamma)_{a} = |\vec{p}_{a}|/m_a$ is the boost-velocity factor. We are interested in decays that occur before the detector, a distance $L$ away: $0\leqslant h\leqslant L$. In the case of DUNE, $L=574$~m.  
%
%

In the ALP rest-frame, the two HNLs are produced back-to-back and the decay is, of course, isotropic. In order to compute the HNL flux, we simulate ALP decays in the lab frame and assume the detector to be a cylinder with radius $r$, aligned with the beam direction (``on-axis''). Since the HNLs are produced at a distance $h$ from the target (see Fig.~\ref{fig:Schematic}), HNLs reach the detector as long as their production angles (relative to the beam direction) in the lab frame are smaller than the opening angle of the detector $\theta_{\rm det}=\arctan(r/L-h)$. 

The total flux of HNLs arriving at the detector can be computed by combining the differential ALP flux and the decay rate. It is
\begin{align}\label{eq:NDflux}
    \phi_{N}&= \int \sin\theta_a\frac{d^2\phi_{a}}{dE_{a}d\theta_a}\frac{{\rm d}P^{a}_{\rm{decay}}}{{\rm d}h}dE_{a}d\theta_a\nonumber
    \\
    &\times\Theta(\theta_\text{det} - \theta)P^{N}_{\rm{sur}}\frac{d\Omega}{4\pi}dh
\end{align}
where $d\Omega$ ad $\theta$ refer to the direction of the HNL and $P^{N}_{\rm{sur}}$ is the probability that an HNL reaches the detector before decaying, 
\begin{equation}
P^{N}_{\rm{sur}}=e^{-\Gamma_N(L-h)/(\beta\gamma)_{N}}.
\end{equation}
$\Gamma_N,(\beta\gamma)_N$ are, respectively, the decay width and boost-velocity factor of the HNL. 

\begin{figure}[ht]
    \centering
\includegraphics[width=0.47\textwidth]{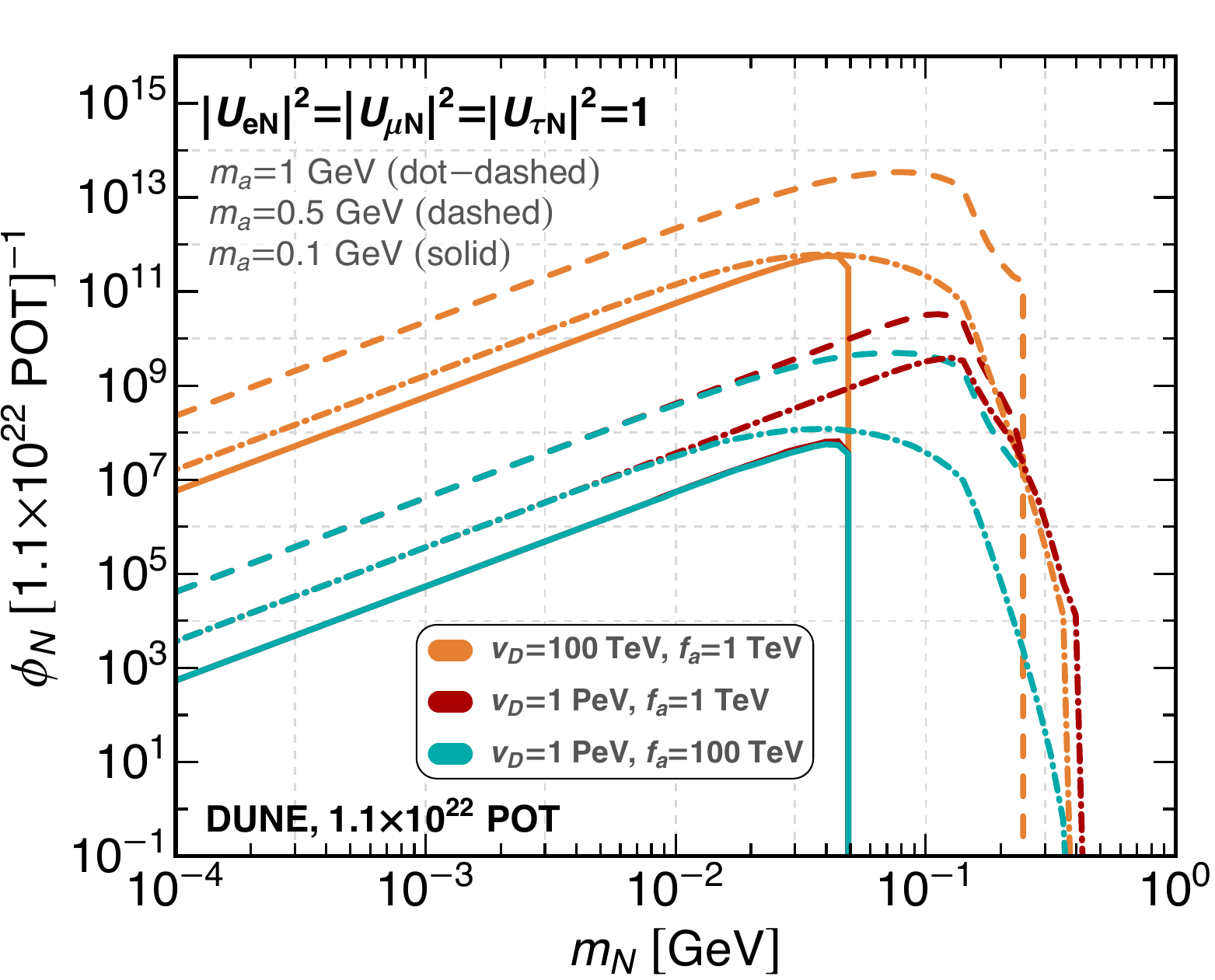}
\caption{HNL flux at the detector as a function of the HNL mass $m_N$, assuming a total exposure of $1.1\times10^{22}$ protons on target. Different colors correspond to different values of $f_{a}$ and $v_D$ (labelled), while the solid (dashed) curves correspond to $m_{a}=0.2~(0.8)~$GeV.  
  }
\label{fig:HNLFlux}
\end{figure}

Fig.~\ref{fig:HNLFlux} depicts the HNL flux at the detector as a function of the HNL mass $m_{N}$, for different values of $f_a$ and $v_D$, for $m_a=0.2~(0.8)~$GeV [solid (dashed) curves] and $|U_{\alpha 4}|^2$ equals one for all $\alpha=e,\mu,\tau$. The HNL flux gets enhanced for $m_a=0.5$ due to the resonance-like behavior in Fig 2. The ALP production rate is proportional to $1/f_a^2$ while the ALP decay rate into HNLs is proportional to $(f_a/v_D^2)^2 \exp{-(f_a m_a m_N /v_D^2)^2 h/|p_a|}$. For our choices of parameters, ALP always decays between the source and the detector.   

{\bf HNL event rates.} HNLs decay via SM interactions through their mixing with the active neutrinos; we assume the dark sector parameters are such that there are no other allowed decay modes. The total and partial decay widths are governed by the HNL mass, $m_N$, and the elements of the mixing matrix $U_{\alpha 4}$, $\alpha=e,\mu,\tau$~\cite{Berryman:2017twh,Pal:1981rm,Gorbunov:2007ak,Ballett:2019bgd}. HNL decays inside the detector are potentially observable and constitute our signal.   

For a given exposure time $\mathcal{T}$, the total event rate from HNL decays into a final state $X$, $N_{\rm signal}(X)$, at the detector is given by a convolution of the decay probability of the HNL into $X$ and the HNL flux, Eq.~(\ref{eq:NDflux}): 
\begin{align}\label{eq:signalrate}
    N_{\rm signal}(X)&=\mathcal{T} \int \sin\theta_a\frac{d^2\phi_{a}}{dE_{a}d\theta_a}\frac{{\rm d}P^{a}_{\rm{decay}}}{{\rm d}h}dE_{a}d\theta_a\nonumber
    \\
    &\times\Theta(\theta_\text{det} - \theta)P^{N}_{\rm{sur}}\frac{d\Omega}{4\pi}dh  \nonumber
        \\&\times P^{N}_{\rm{decay}}\, \mathbf{Br}(N\to X)
\end{align}
%
where $\mathbf{Br}(N\to X)$ is the branching ratio of the HNL decay into $X$ and the probability the decay happens inside the volume of the detector, with the length $\Delta L$, is 
\begin{equation}
    P^N_{\rm decay}= e^{-\Gamma_N^{\rm{Total}} (L-h)/(\beta\gamma)_{N}} \left[ 1 - e^{-\Gamma_N^{\rm{Total}} \Delta L/(\beta\gamma)_{N} } \right]\,.
\end{equation}
The HNL lifetimes and the various branching fractions can be found in Appendix~A of ref.~\cite{Berryman:2017twh}; we depict different branching ratios as a function of $m_N$ in Fig.~\ref{fig:BrND}, in the supplemental materials. We consider the charged-current mediated channels $X=e^\pm\pi^\mp,\mu^\pm\pi^\mp,\nu e^\pm\mu^\mp$, the neutral-current mediated channel $X\nu\pi^0$, the ``mixed'' channels $X=\nu \ell^+ \ell^-$ ($\ell=e,\mu$), and the one-loop channel $X=\nu\gamma$. We pay special attention to $X=\nu e^+ e^-$ since it has the largest branching fraction for $m_N$ below $m_\pi$ and contributes appreciably for larger masses. At small $m_N$, $X=\nu\gamma$ is also of interest, even though it has a much smaller branching smaller. 


{\bf DUNE Analysis and Results.} 
\label{sec:AnalysisResults}
In order to estimate the sensitivity of the DUNE near-detector facility to HNLs produced via ALP decays, we assume a cylindrical (radius $r=2.6$~m and length $\Delta L=10$~m), on-axis, $50$~ton liquid argon near-detector, located $L=574$~m from the target. We assume ten years of data taking with an exposure of $1.1\times 10^{21}$~POT per year~\cite{DUNE:2020ypp}. We estimate the sensitivity to the HNL parameter space using a naive $\chi^2$ analysis based on the expected HNL rate $N_{\rm{HNL}}$, where for  $90\%$~CL and 2-degrees of freedom we use  $N_{\rm{HNL}}=4.61$. The model of interest contains eight, assumed-to-be-uncorrelated, parameters: $(f_{a}, v_D, m_{a}, m_{N}, |U_{\alpha 4}|^2)$, $\alpha=e,\mu,\tau$.  For each final state, we perform an independent analysis and ``turn on'' only one $|U_{\alpha 4}|^2$ at a time (i.e., only one among $|U_{e N}|^2$, $|U_{\mu N}|^2$, and $|U_{\tau N}|^2$ are nonzero at a given analysis). Unless otherwise noted, we fix $m_{\pi_D}=3m_{N}$, for concreteness. 

For simplicity, we do not include backgrounds in our analyses. We expect this to be a reasonable assumption for some HNL-decay final states, including $e^+e^-\nu$, but are quite confident this is not the case for other final states, where significant beam-induced backgrounds are expected. A dedicated study is necessary for properly estimating the different backgrounds. We leave such background studies for future work.

\begin{figure}[ht]
    \centering
    \includegraphics[width=0.44\textwidth]{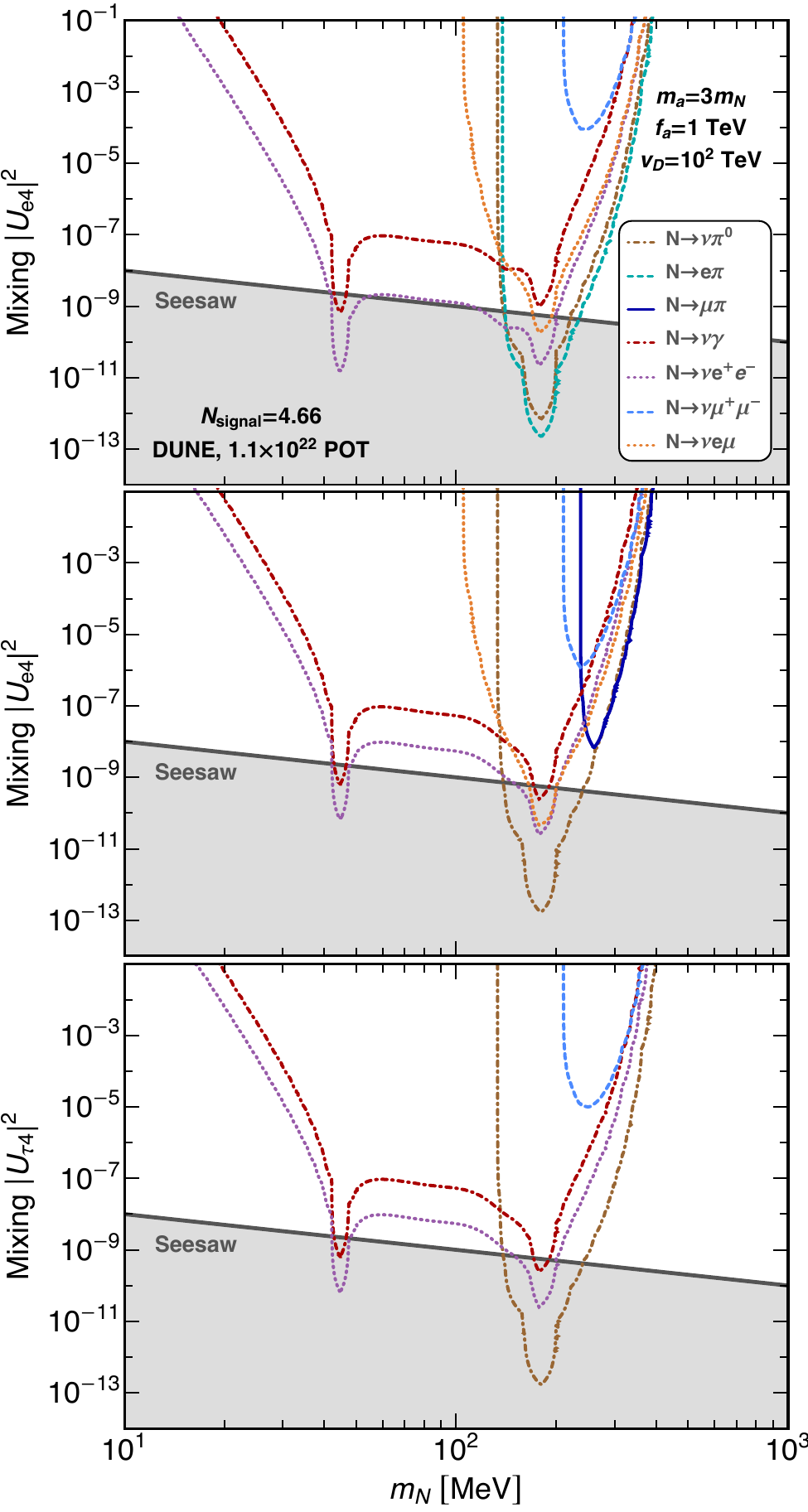}
    \caption{The sensitivity of the DUNE near detector to the HNLs produced from the dark pion decay, assuming a total of $1.1\times10^{22}$ protons-on-target (POT). The panels correspond to different HNL mixing structures, namely mixing only with $\nu_e$ (top panel), $\nu_\mu$ (middle panel) and $\nu_\tau$ (bottom panel), respectively. 
    }
    \label{fig:AllowedRegions}
\end{figure}

Our results in the $m_{N}\times |U_{\alpha 4}|^2$--plane, for fixed $f_{a}=1~$TeV and $v_D=10^2~$TeV are depicted in Fig.~\ref{fig:AllowedRegions}. The top, middle, and bottom panels correspond $\nu_e$-coupled, $\nu_\mu$-coupled, and $\nu_\tau$-coupled HNLs, respectively. For very small $|U_{\alpha 4}|$, HNL production is severely suppressed since the HNLs are too long-lived to decay inside the detector. 
There are regions of enhanced sensitivy  at $m_{N}\sim40~$MeV and $m_{N}\sim200~$MeV. These correspond to the resonant-production peaks observed in Fig.~\ref{fig:DarkPionFlux} when $m_a\sim 120$~MeV and $\sim 600$~MeV, respectively. For $m_N$ smaller than, roughly, the pion mass, only the $\nu e^+e^-$ and $\nu\gamma$ channels are open. These remain competitive for larger values of $m_N$ but are superceded by decay modes with larger branching ratios. Results for $v_D=10^3$~TeV are depicted in Fig.~\ref{fig:AllowedRegions2}, in the supplemental materials.   

Regardless of whether the HNL is purely $\nu_e$-coupled (only $|U_{e4}|\neq 0$), $\nu_\mu$-coupled (only $|U_{\mu4}|\neq 0$), or $\nu_\tau$-coupled (only $|U_{\tau4}|\neq 0$), the overall sensitivities are approximately the same. This is to be contrasted with the standard scenario in which the HNLs are the product of charged-meson decays. There, due to the comparative sizes of the respective parent meson fluxes, the sensitivity to $\nu_\mu$-coupled HNLs is superior to that of $\nu_e$-coupled HNLs, which is in turn superior to that of $\nu_\tau$-coupled HNLs. This is especially relevant for $\tau$-coupled HNLs whose production is severely suppressed by the flux of charm mesons (e.g. $D_s$) at the target, or, in the case of lower-energy facilities, completely absent. The enhanced production of $\tau$-coupled HNLs makes it possible to better probe the $|U_{\tau 4}|$ coupling, as well as search for the potential detector signatures of the $\tau$-lepton~\cite{Conrad:2010mh,DeGouvea:2019kea,Machado:2020yxl,Dev:2023rqb,Meighen-Berger:2023xpr,MammenAbraham:2022xoc}.

For the ALP parameters of choice, in this work we are sensitive to significantly smaller $|U_{\alpha 4}|$ values. This is expected. For charged-meson-produced HNLs, the signal rate is naively proportional to $|U_{\alpha 4}|^4$, while for neutral-meson-produced HNLs, the signal rate is naively proportional to $|U_{\alpha 4}|^2$. For the scenario considered here, we find that DUNE is sensitive to HNLs whose parameters naturally agree with expectation from the naive type-I seesaw. These regions of parameter space are highlighted in grey in Fig.~\ref{fig:AllowedRegions} and correspond to $|U_{\alpha 4}|^2\le 0.1~{\rm eV}/m_N$~\cite{Ballett:2019bgd}.


\medskip 
{\bf Conclusion.} 
\label{sec:Conclusion}
SM-singlet fermions are often among the ingredients behind nonzero neutrino masses. If this is the case, and if these new particles are light enough, they can be produced and detected in fixed-target-like experiments, like the DUNE near-detector facility. Here, we explore the fact that the production and detection of SM-singlet fermions is model dependent. We show that if these belong to a richer -- but rather generic -- dark sector, it is possible that an experiment like DUNE can explore regions of parameter space relevant for the neutrino mass puzzle.

In more detail, After electroweak symmetry breaking, SM-singlet fermions, also referred to as heavy neutral leptons (HNL), naturally mix with SM neutrinos. If these are light enough, they can be produced via the weak interactions in the decay of charged mesons (e.g., $\pi^+\to\mu^+ N$) and decay via the weak interactions into leptons and other light states. On the other hand, axion-like particles (ALP) will naturally decay, often with large branching ratio, into HNLs if these belong to the same dark sector. Since ALPs mix with the SM neutral pseudoscalar hadrons, we explored for the first time the consequences of the fact that the HNL flux at a fixed-target-like facility might be predominantly associated to the production of neutral mesons, including the $\pi^0$ and the $\eta$. If this is the case, the physics responsible for HNL production and decay is not directly related and the production rate is not suppressed by the active--sterile mixing parameters. Furthermore, experiments like DUNE might be sensitive to HNLs that are too weakly coupled, directly, to the SM to be produced in the decay of charged mesons. This is naively expected, for example, of HNLs that play a direct part in the type-I seesaw mechanism.  



Our analyses are confined to the DUNE experiment but we anticipate that the HNL--ALP connection will lead to interesting consequences across the spectrum of neutrino facilities, including ICARUS, SBND, $\mu$BooNE, COHERENT, CCM, and FASER$\nu$ and provide a new avenue for future research and experimentation in the quest to uncover the mysteries of HNLs and their role in particle physics.

\begin{acknowledgments}
The work of AdG is supported in part by the U.S.~Department of Energy grant DE-SC0010143 and in part by the NSF grant PHY-1630782. The work of BD is supported  by the U.S.~Department of Energy Grant DE-SC0010813. The work of ZT is supported by the Neutrino Theory Network Program Grant No. DE-AC02-07CHI11359 and the U.S.~Department of Energy under the award number DE-SC0020250. She also appreciate the hospitality of Aspen Physics Center where this work was partially developed. The work of IMS is supported by the U.S. Department of Energy grant DE-SC0020262. AA acknowledges partial support from the European Union’s Horizon 2020 Research and Innovation Programme under the Marie Sklodowska-Curie grant agreement no. 860881-HIDDeN and the Marie Sklodowska-Curie Staff Exchange grant agreement no. 101086085–ASYMMETRY; from Fermi National Accelerator Laboratory, managed and operated by Fermi Research Alliance, LLC under Contract No. DE-AC02-07CH11359 with the U.S. Department of Energy; and from the Spanish Research Agency (Agencia Estatal de Investigación) through the grant IFT Centro de Excelencia Severo Ochoa no. CEX2020-001007-S and the grant RYC2018-024240-I, funded by MCIN/AEI/10.13039/501100011033 and by “ESF Investing in your future”. 
\end{acknowledgments}

\vspace{8cm}
\appendix
\section{Supplemental Material}
\label{app:AllEq}

\begin{figure}
    \centering
\includegraphics[width=0.43\textwidth]{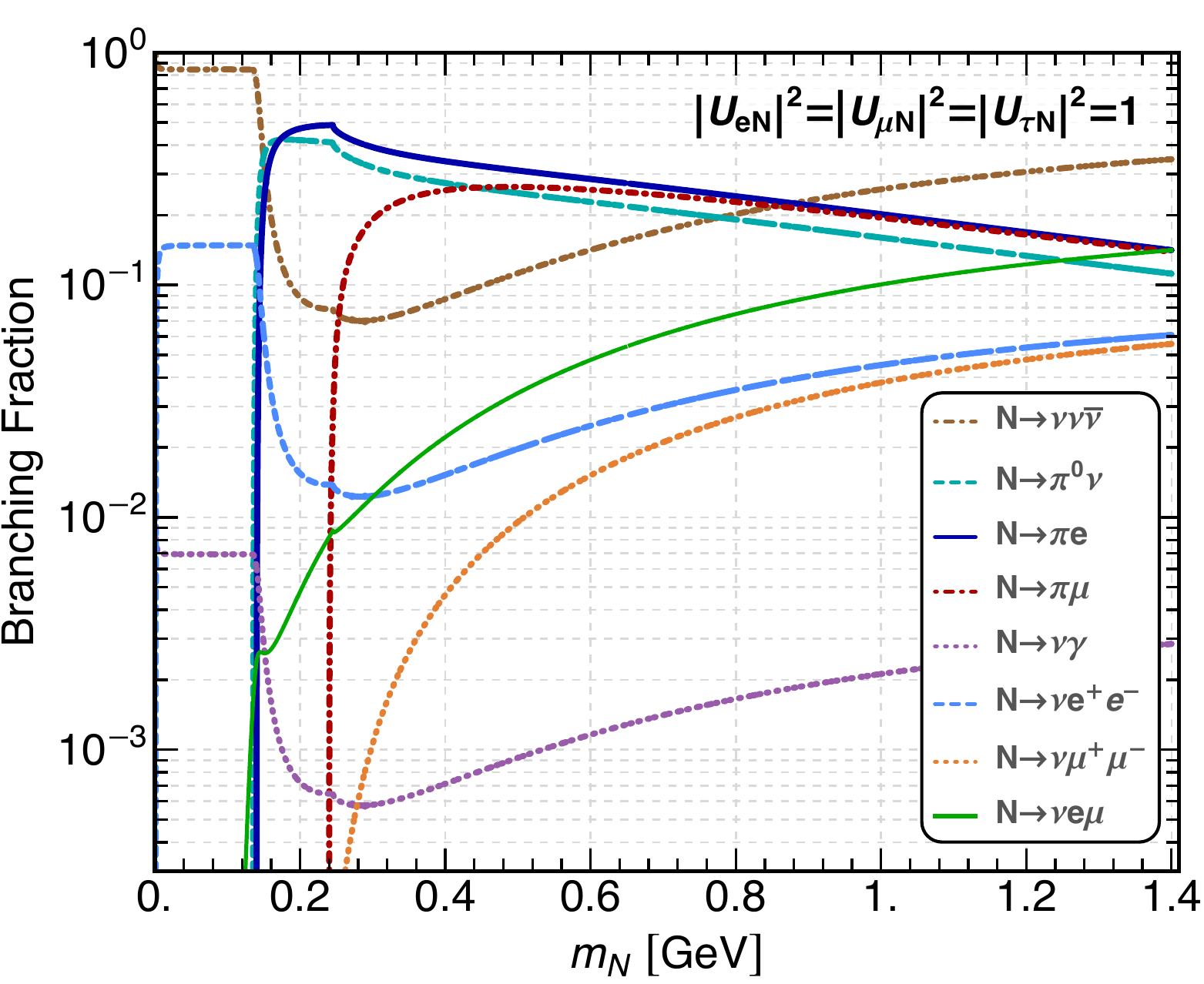}
\caption{Branching fractions for different decay modes of the heavy neutral lepton (HNL), as a function of the HNL mass $m_N$. 
  }
\label{fig:BrND}
\end{figure}

For the expressions of the partial HNL decay widths we have used the expressions in appendix A of~\cite{Berryman:2017twh}, as well as Ref.~\cite{Pal:1981rm,Gorbunov:2007ak,Ballett:2019bgd}. In general with all the decay channels that are kinetically allowed for the mass range we consider in this work, the total decay width reads:
\begin{align}
    \Gamma_N^{\rm{Total}} = \,\,
    &2\sum_{\alpha=e,\mu,\tau}\Big[ \Gamma(N\to \nu_\alpha \bar\nu \nu) +\Gamma(N\to\nu_\alpha \gamma) 
    \nonumber\\
    &+\Gamma(N\to\nu_\alpha \pi^0) \Big]+2\sum_{\alpha=e,\mu}\Big[\Gamma(N\to\pi^+\ell_\alpha) 
    \nonumber\\
    &+ \Gamma(N\to\ell^-_\alpha\ell^+\nu)
    +\Gamma(N\to\ell^-_\alpha\ell^+\nu)
    \nonumber\\
    &+\Gamma(N\to\nu_\alpha\ell^-\ell^+)\Big]\,,
\end{align}
where the overall factor of 2 accounts for the fact that the HNLs are Majorana fermions. We have shown the branching fractions in Fig.~\ref{fig:BrND}, where we have assumed all the mixing angles $|U_{\alpha 4}|^2$ are equal, and can be cancelled from the expressions of the branching fractions.

\begin{figure}
    \centering
    \includegraphics[width=0.4\textwidth]{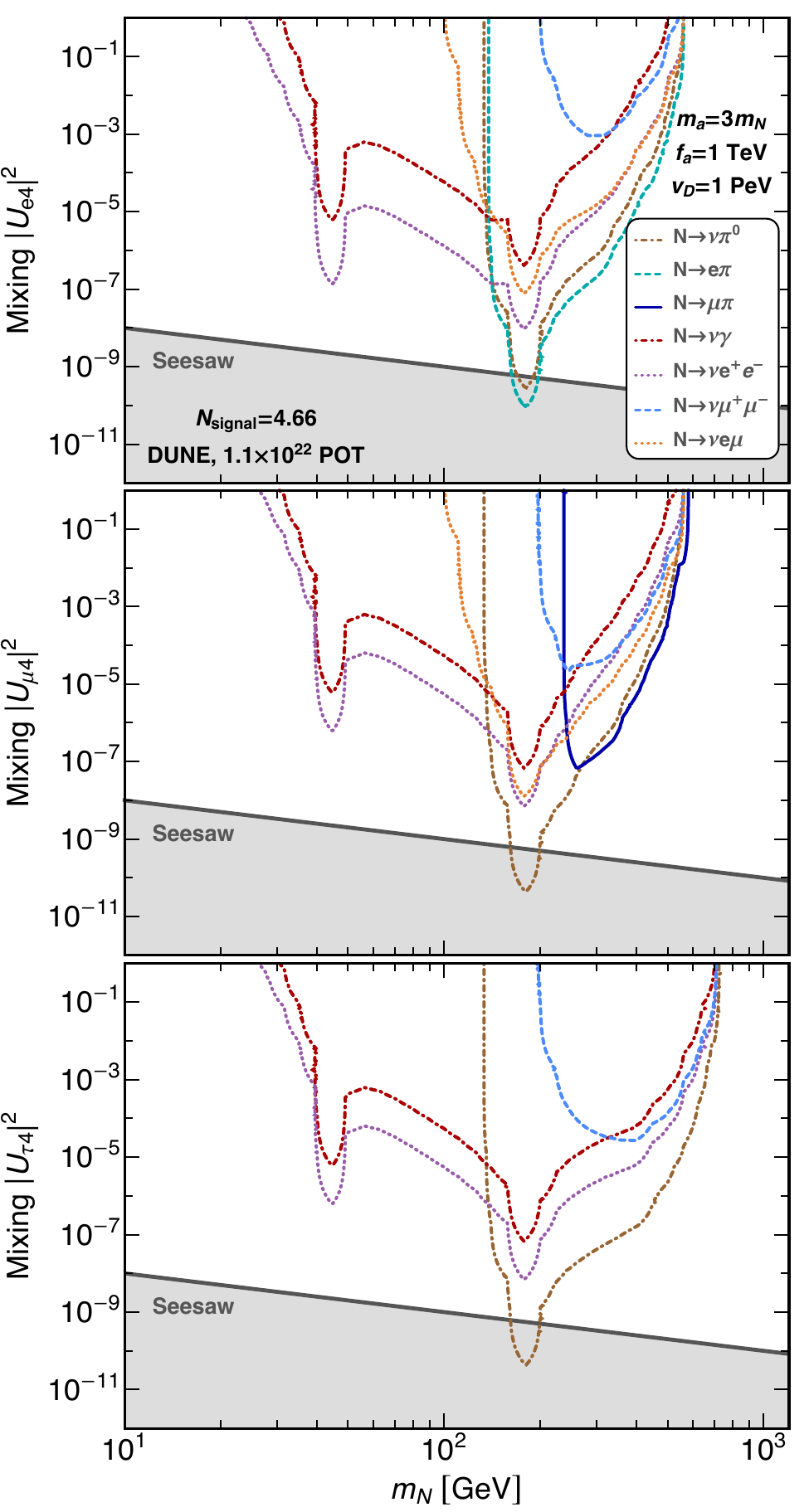}
    \caption{Same as Fig.~\ref{fig:AllowedRegions}, for $v_D=10^3~$TeV. 
    }
    \label{fig:AllowedRegions2}
\end{figure}

To demonstrate the dependence of our final results on the value of the dark vev $v_D$, we have made Fig.~\ref{fig:AllowedRegions2}, for which all the parameters are the same as in Fig.~\ref{fig:AllowedRegions}, but $v_D=10^3~$TeV. 

\newpage
\bibliographystyle{apsrev4-1}
\bibliography{main}{}

\end{document}